\documentstyle[aps]{revtex}
\newcommand{\be}{\begin{equation}}
\newcommand{\ee}{\end{equation}}
\newcommand{\ba}{\begin{eqnarray}}
\newcommand{\ea}{\end{eqnarray}}
\newcommand{\bas}{\begin{eqnarray*}}
\newcommand{\eas}{\end{eqnarray*}}

\newcommand{\bC}{\mbox{{\bf C}}}
\newcommand{\bZ}{\mbox{{\bf Z}}}
\newcommand{\bZp}{\mbox{{\bf Z}}_{\geq 0}}

\begin{document}
\draft
%
%
%
%
\title{%
Jack polynomials with prescribed symmetry
\\
and hole propagator 
of spin Calogero-Sutherland model\\
} 
\author{%
Yusuke Kato\cite{permanent} 
and Takashi Yamamoto\cite{crest}}
\address{%
Institut f\"ur Theoretische Physik, 
Universit\"at zu K\"oln, 
Z\"ulpicher str.77, 
D-50937, 
K\"oln, 
Germany\\
and\\
Department of Physics, 
Tohoku University, 
Sendai 980-77, 
Japan}

\maketitle
\begin{abstract}
We study the hole propagator of the Calogero-Sutherland model 
with SU(2) internal symmetry. 
We obtain the exact expression 
for arbitrary non-negative integer coupling parameter $\beta$ 
and prove the conjecture proposed by one of the authors. 
Our method is based on the theory of the Jack polynomials with
a prescribed symmetry. 
\end{abstract}
\pacs{71.10.Pm, 05.30.-d}
%
%
%

\section{Introduction}
\label{intro}
One of the goals in physics of interacting particle systems 
is to understand the dynamics. 
In particular, dynamical properties in one-dimensional systems 
are anticipated to be intriguing, because interaction effects 
are crucial and naive pictures 
based on perturbations lose their validity. 
As a non-perturbative approach, 
conformal field theory is a powerful method 
to study the low energy physics of 
the Tomonaga-Luttinger liquid. 
Beyond the conformal limit, 
on the other hand, 
integrable systems give us opportunities 
for analytical study of the dynamics. 
Among them, 
the Calogero-Sutherland (CS) model \cite{Cal,Suth} of  
particles interacting 
with the two-body inverse square interaction
provides the simplest example of 
interacting particle systems with nontrivial dynamics. 
For the spinless CS model, 
the density-density correlation function \cite{SLA,Ha,LPS,MP}, 
hole propagator \cite{Ha,LPS,HZ} 
and particle propagator \cite{ZH,SLP} 
have been obtained analytically. 

The spinless CS model has a number of variants. 
One of them is the spin CS model 
\cite{HaHaldane,MP2}. 
This model describes $n$ particles with coordinates 
$X=(X_1,\dots,X_n)$
moving along a circle of length $L$ and 
each particle possesses a spin with $p$ possible values.
The Hamiltonian of the model is given by
\begin{equation}
\label{hamiltonian}
\hat{H}_n
=
-\sum_{i=1}^n \frac{\partial^2}{\partial X_i^2}
+
2\left(\frac{\pi}{L}\right)^2\sum_{1\leq i<j\leq n}
\frac{\beta(\beta + P_{ij})}{\sin^2\frac{\pi}{L}(X_i - X_j)},
\end{equation}
where $\beta$ is a coupling parameter
and $P_{ij}$ is the spin exchange operator.
In this paper, 
we take $\beta$ to be a non-negative integer.

The spin CS model with $p=2$ is particularly relevant to 
the condensed matter physics; 
most one-dimensional systems are realized 
experimentally in electron 
systems and hence we should take account of spin degrees 
of freedom of
each particle. Further we can regard the spin CS model 
with $p=2$ as a one-dimensional variant 
of singlet fractional quantum Hall (FQH) system; 
the ground state of (\ref{hamiltonian}) 
can be derived from the Halperin 
wavefunction \cite{Halperin} for the singlet FQH state 
by the restriction of particle coordinates 
on a ring in two-dimensional plane.     

For the dynamics of the spin CS model with $p=2$, 
the expression for the hole part of Green function 
has been proposed \cite{Green} 
relying on the finite size calculations. 
Subsequently the dynamical density-density 
and spin-spin correlation functions have been derived 
by Uglov \cite{Uglov} in an exact treatment. 
Since the paper \cite{Uglov} appeared, however, 
exact derivations of the hole propagator have 
been still missing.
Recently, in \cite{Dunkl3},
Dunkl has developed the theory 
of the Jack polynomials with a prescribed symmetry; 
those polynomials 
are symmetric or alternating 
with respect to the interchange
of certain subsets of variables.
Dunkl's results allow us to derive `the binomial formula', 
which is directly related to the matrix element 
of the local field operator in the spin CS model. 

The aim of this paper is to prove the earlier conjecture 
on the hole propagator \cite{Green} 
utilizing Dunkl's results \cite{Dunkl3}. 
Though the method in this paper is applicable also 
to the general SU($p$) case, 
we concentrate on the SU(2) case for simplicity. 

Here we shall recall the conjecture in \cite{Green} 
on the hole propagator. 
In the thermodynamic limit,
the expression for the hole propagator $G(r,t)$
is expected to be
\begin{equation}
\label{conjecture}
G(r,t)
=
c(\beta) 
\prod_{k=1}^\beta \int_{-1}^{1}{\rm d}u_{k}
\prod_{l=1}^{\beta+1}\int_{-1}^{1}{\rm d}v_{l}
\left|F(u,v)\right|^2 
\exp\left[-{\rm i}
\left(E(u,v)t
-Q(u,v)r\right)\right]\label{holetherm}. 
\end{equation}
Here $c(\beta)$ is a constant factor
and $u=(u_1,\cdots,u_{\beta})$, $v=(v_1,\cdots,v_{\beta+1})$
represent normalized velocities of the quasiholes.
In the expression (\ref{holetherm}), 
$F$, $Q$ and $E$ 
represent the form factor, momentum and energy, respectively. 
The explicit forms of them are as follows: 
The form factor $F$ 
is given by 
\be
\label{form-factor-tdlim}
F(u,v)=
\frac{%
\prod_{1\leq k<l\leq\beta}
\left(u_{k} -u_{l}\right)^{g_{\rm d}}
\prod_{1\leq k<l\leq\beta+1}
\left(v_{k} -v_{l}\right)^{g_{\rm d}}
\prod_{k=1}^{\beta}
\prod_{l=1}^{\beta+1}
\left(u_{k}-v_{l}\right)^{g_{\rm o }}}
{%
\prod_{k=1}^{\beta}\left(1-u_{k}^2\right)^{(1-g_{\rm d})/2}
\prod_{l=1}^{\beta+1}\left(1-v_{l}^2\right)^{(1-g_{\rm d})/2}},
\ee
where
$g_{\rm d}=\left(\beta+1\right)/(2\beta+1)$ 
and $g_{\rm o}=-\beta /(2\beta+1)$.
The momentum $Q$ and energy $E$ are respectively given by
\ba
\label{momentum-tdlim}
Q(u,v)
&=&
\frac{\pi\rho_0}{2}
\left(\sum_{k=1}^{\beta}u_{k}
+\sum_{l=1}^{\beta+1}v_{l}\right),
\\
\label{energy-tdlim}
E(u,v)
&=&
-\left(2\beta+1\right)\left(\frac{\pi \rho_0}{2}\right)^2
\left(\sum_{k=1}^{\beta}u_{k}^2 
+\sum_{l=1}^{\beta+1}v_{l}^2\right). 
\ea
Here $\rho_0=2M/L$ is the mean density of particles.
In what follows, 
we prove this conjecture and show that the constant $c(\beta)$ 
is given by 
\be
\label{conjecture-coeff}
c(\beta)
=
\frac{\rho_0}
     {4\left(2\beta+1\right)^\beta\Gamma\left(\beta+2\right)}
\prod_{k=1}^{2\beta +1}
\frac{\Gamma\left(\left(\beta+1\right)/(2\beta+1)\right)}
     {\Gamma\left(k/(2\beta+1)\right)^2},
\ee
where $\Gamma(z)$ is the gamma function.
The physical implication of the expressions 
(\ref{conjecture}), (\ref{form-factor-tdlim}),
(\ref{momentum-tdlim}) and (\ref{energy-tdlim}) 
has been discussed in \cite{Green,KYA}.  

This paper is organized as follows. 
In next section, 
we define the Jack polynomials with a prescribed symmetry 
and discuss their basis properties. 
Particularly, we derive the binomial formula 
of the Jack polynomials with the prescribed symmetry 
with the use of Dunkl's results.  
In section ref{propagator}, 
we obtain the hole propagator for a finite number of particles 
using the mathematical formulae presented 
in section \ref{math-pre}
and present the derivation of the expression 
(\ref{conjecture}).  

\section{Jack polynomials with prescribed symmetry}
\label{math-pre}
In this section, 
we present mathematical results necessary 
to derive the expression for the hole propagator. 
First, as a preliminary, we fix our notations. 
Second, we define the non-symmetric Jack polynomials. 
Third, we define the Jack polynomials 
with a prescribed symmetry 
in terms of the non-symmetric Jack polynomials. 
Fourth, we present the basic formulae of 
the Jack polynomials with the prescribed symmetry; the norm, 
Cauchy formula, and evaluation formula are discussed. 
Last, from the Cauchy and evaluation formulae, 
we derive the binomial formula, 
which gives the matrix element of the local field operator 
in the calculation of the hole propagator. 
\subsection{Notations}
\label{notation}
First of all, we fix notations 
(see Refs. \cite{Macd,KS,Sahi}).
For a fixed non-negative integer $n$, let 
$\Lambda_n
=\{\eta=(\eta_1,\eta_2,\cdots,\eta_n)\,
|\,\eta_i\in\bZp,\,1\leq i\leq n\}$
be the set of all compositions 
with length less than or equal to $n$.
The weight $|\eta|$ 
of a composition 
$\eta=(\eta_1,\eta_2,\cdots,\eta_n)\in\Lambda_n$
is defined by $|\eta |=\sum_{i=1}^n\eta_i$.
The length $l(\eta)$ of $\eta$ is defined 
as the number of non-zero $\eta_i$ in $\eta$. 
The set of all partitions with length less than or equal to $n$
is defined by 
$\Lambda_n^+
=\{\lambda=(\lambda_1,\lambda_2,\cdots,\lambda_n)\in\Lambda_n\,
|\,\lambda_1\geq\lambda_2\geq\cdots\geq \lambda_n\geq 0\}$. 
The dominance order $<$ on partitions
is defined as follows: 
for $\lambda$, $\mu\in \Lambda^+_n$,
$\lambda\le \mu $ if $\vert\lambda\vert=\vert\mu\vert$ 
and $\sum_{i=1}^k \lambda_i\le  \sum_{i=1}^k \mu_i$ 
for all $k=1,\cdots,n$.   
For a composition $\eta\in\Lambda_n$, 
we denote by $\eta^+$ the (unique) partition 
which is a rearrangement of the composition $\eta$. 
Now we define a partial order $\prec$ 
on compositions as follows: 
for $\nu,\eta\in\Lambda_n$, $\nu \prec\eta$ 
if $\nu^+ <\eta^+$ with dominance ordering 
on partitions
or if $\nu^+ =\eta^+$ 
and $\sum_{i=1}^k \nu_i \le \sum_{i=1}^k \eta_i$ 
for all $k=1,\cdots,n$. 

For a given composition
$\eta=(\eta_1,\eta_2,\cdots,\eta_n)\in\Lambda_n$
and 
pairs of integers $s=(i,j)$ satisfying $1\leq i\leq l(\eta)$ 
and $1\leq j\leq \eta_i$, we define the following quantities:
\ba
&&\label{arm}
a(s)=\eta_i-j,
\\
&&\label{co-arm}
a'(s)=j-1, 
\\
&&\label{leg}
l(s)=\#\{k\in\{1,\cdots,i-1\}|j\leq \eta_k+1\leq\eta_i\}
    +\#\{k\in\{i+1,\cdots,n\}|j\leq \eta_k\leq\eta_i\}, 
\\
&&\label{co-leg}
l'(s)=\#\{k\in\{1,\cdots,i-1\}|\eta_k\geq\eta_i\}
     +\#\{k\in\{i+1,\cdots,n\}|\eta_k>\eta_i\}.
\ea
Here, for a set $A$, $\#A$ denotes the number of elements.
In the above expressions, $a(s)$, $a'(s)$, $l(s)$ and $l'(s)$ 
are called
arm-length, coarm-length, leg-length, and coleg-length,
respectively.
Further, for a composition $\eta\in\Lambda_n$
and a parameter $\beta$, 
we define the following four quantities:
\ba
\label{d}
&& d_\eta=\prod_{s\in\eta}((a(s)+1)/\beta+l(s)+1), \\
\label{d'}
&& d'_\eta=\prod_{s\in\eta}((a(s)+1)/\beta+l(s)), \\
\label{e}
&& e_\eta=\prod_{s\in\eta}((a'(s)+1)/\beta+n-l'(s)), \\
\label{e'}
&& e'_\eta=\prod_{s\in\eta}((a'(s)+1)/\beta+n-l'(s)-1).
\ea 
\subsection{Non-symmetric Jack polynomials}
\label{nonsymm-jack}
Now we define the non-symmetric Jack 
polynomials \cite{Opdam,Macdonald}. 
For this purpose, we define the Cherednik-Dunkl operators 
\cite{Dunkl0,Cherednik} as
\be
\label{dunkl}
\hat{d}_i
=
x_i\frac{\partial}{\partial x_i}
+
\beta\sum_{k=1}^{i-1}\frac{x_i}{x_i-x_k}(1-s_{ik})
+
\beta\sum_{k=i+1}^{n}\frac{x_k}{x_i-x_k}(1-s_{ik})
+
\beta(1-i)
\ee
for $1\leq i\leq n$. The operator 
$s_{ij}=(i,j)$ is the transposition which 
interchanges coordinates $x_i$ and $x_j$
($s_{ij}$ is called the coordinate exchange operator 
\cite{Polychronakos}).
The operator $\hat d_i$ 
is a mapping in homogeneous polynomials of
$x=(x_1,\cdots,x_n)$. Further all the operators 
$\{\hat d_i\}$ commute with each other, 
and hence these operators can be diagonalized simultaneously. 

For a given composition 
$\eta$, 
the non-symmetric Jack polynomial
$E_\eta(x;1/\beta) $ is defined as 
the $|\eta|$-th order homogeneous polynomial 
satisfying the following two conditions:
\begin{enumerate}
\item 
The polynomial $E_\eta(x;1/\beta)$ has the form of
\be
\label{non-symm-Jack}
E_\eta(x;1/\beta)
=
x^\eta
+
\sum_{\nu\in \Lambda_n
       \atop
     {\scriptscriptstyle \nu\prec\eta}}
c_{\nu \eta}x^\nu,  
\ee
in terms of the monomials 
$x^\nu=x_1^{\nu_1}\cdots x_n^{\nu_n}$. 
\item 
$E_\eta(x;1/\beta) $ 
is a simultaneous eigenfunction of $\hat d_i$ 
for $1\leq i \leq n$. 
\end{enumerate}

Here we remark on two important properties 
of the non-symmetric Jack polynomial $E_\eta$. 
One is the eigenvalue of $E_\eta$ for $\hat d_i$, 
which is given by
\be
\label{bareta}
\bar{\eta}_i
=
\eta_i
-\beta
\left(
\#\{k\in\{1,\cdots,i-1\}|\eta_k\geq\eta_i\}
+
\#\{k\in\{i+1,\cdots,n\}|\eta_k>\eta_i\}
\right).
\ee
The other is the action of the transposition 
$s_i:=s_{ii+1}=(i,i+1)$ 
on $E_\eta$ \cite{KS,Sahi},
\begin{equation}
\label{trans-property}
s_iE_\eta
=
\left\{
\begin{array}{ll}
\xi_i E_\eta
+
\left(1-\xi_i ^2\right)
E_{s_i\eta},
&\quad
\eta_i>\eta_{i+1},
\\
E_\eta,
&\quad
\eta_i=\eta_{i+1},
\\
\xi_i E_\eta
+
E_{s_i\eta},
&\quad
\eta_i<\eta_{i+1}, 
\end{array}
\right.
\end{equation}
where 
$\xi_i 
=\beta/\left(\bar \eta_i -\bar \eta_{i+1}\right)$.
In particular, the property (\ref{trans-property}) 
plays an important role in the proof of 
the basic properties 
of the Jack polynomials with a prescribed symmetry. 
\subsection{Jack polynomials with prescribed symmetry}
\label{jack-with-symm}
Next we define the Jack polynomials 
with a prescribed symmetry. For this purpose, 
we introduce some notations.
The interval $I=[1,n]$ denotes $\{i\in\bZ|1\leq i\leq n\}$ 
for a positive integer $n$. 
For an integer $m\in I$, 
we define $I_\downarrow=[1,m]$ and $I_\uparrow=[m+1,n]$. 
In addition, we introduce the notations 
$n_\downarrow=\#I_\downarrow(=m)$ 
and 
$n_\uparrow=\#I_\uparrow(=n-m)$.
We consider the subgroup 
$S_{n_\downarrow}\times S_{n_\uparrow}$ of 
the symmetric group $S_n$ which
leaves $\{1,\cdots,n_\downarrow\}$ and 
$\{n_\downarrow+1,\cdots,n\}$ 
invariant.
Further we define 
$\Lambda_n^{\downarrow\uparrow} \subset \Lambda_n$ 
as a set of `partial partitions' 
\begin{equation}
\Lambda_n^{\downarrow \uparrow}=
\{
\mu
=(\mu^\downarrow,\mu^\uparrow)
=(\mu_1^\downarrow,\cdots,
\mu_{n_\downarrow}^\downarrow,
\mu_1^\uparrow,\cdots,
\mu_{n_\uparrow}^\uparrow)\in\Lambda_n|
\mu_1^\downarrow>\cdots>
\mu_{n_\downarrow}^\downarrow,\, 
\mu_1^\uparrow>\cdots>
\mu_{n_\uparrow}^\uparrow
\}.
\end{equation}

Now we consider polynomials
which are alternating under the action of 
$S_{n_\downarrow}\times S_{n_\uparrow}$ 
\cite{Dunkl3,BF1}.
We define the (alternating)
Jack polynomial with the prescribed symmetry
$K_\mu(x;\beta)$ for $\mu
=
(\mu^\downarrow,\mu^\uparrow)
\in\Lambda_n^{\downarrow\uparrow}$
by the following two conditions:
\begin{enumerate}
\item
The polynomial $K_\mu$ has the form of
\begin{equation}
K_\mu(x;\beta)
=
\sum_{\eta=(\eta^\downarrow,\eta^\uparrow)}
a_\eta E_\eta(x;1/\beta)
\end{equation}
with the normalization $a_\mu=1$. 
Here the sums over $\eta^\downarrow$ and $\eta^\uparrow$ 
are taken on the whole rearrangements of 
$\mu^\downarrow$ and $\mu^\uparrow$, respectively.
\item
Under the action of the transposition $s_i$,
the polynomial $K_\mu(x)$ is transformed as 
$s_iK_\mu(x)=-K_\mu(x)$ 
for $i\in I_\downarrow\setminus\{n_\downarrow\}$ 
or $\in I_\uparrow\setminus\{n\}$. 
(Here, for a set $A$ and its subset $B$,
we denote by $A\setminus B$
the complementary set of $B$ in $A$.) 
\end{enumerate}
{}From the above definition and (\ref{trans-property}), 
we can derive the recursion relation 
for the coefficient $a_\eta$
\be
\label{coeff}
a_{s_i\eta}
=
-
\frac{\bar{\eta}_i-\bar{\eta}_{i+1}-\beta}
     {\bar{\eta}_i-\bar{\eta}_{i+1}}
a_{\eta},
\ee
where $i\in I_\downarrow\setminus\{n_\downarrow\}\,
\mbox{or}\,\in I_\uparrow\setminus\{n\}$. From 
the alternating property of $K_\mu$, 
we can also write as
\begin{equation}
\label{KetaEeta}
\rho_\mu K_\mu(x)
=
\sum_{\sigma\in S_{n_\downarrow}\times S_{n_\uparrow}}
\mbox{sgn}(\sigma)\sigma E_\mu(x),
\end{equation}
where the symbol $\mbox{sgn}(\sigma)$ 
denotes the sign of the permutation $\sigma$.  
The expression for  the factor $\rho_{\mu}$ 
can be obtained from the normalization $a_\mu=1$ 
and the relation (\ref{trans-property}) as
\begin{equation}
\label{rho}
\rho_\mu
=
\prod_{s=\downarrow,\uparrow}
\prod_{i,j\in I_s
       \atop
       {\scriptscriptstyle i<j}}
\frac{\bar{\mu}_i-\bar{\mu}_j-\beta} 
     {\bar{\mu}_i-\bar{\mu}_j} 
\end{equation}
for $\mu \in \Lambda^{\downarrow\uparrow}_n$.
\subsection{Basic properties}
\label{property}
Next we discuss the basic properties of 
the Jack polynomials with the prescribed symmetry. 
The combinatorial norm, 
integral norm, 
Cauchy formula and evaluation formula are discussed. 

\noindent
a) combinatorial norm:
Define the polynomials $\{q_\eta(x)\}_{\eta\in\Lambda_n}$ by
\be
\label{generating-function-non-symm-Jack}
\Omega(x|y)
=
\prod_{i=1}^n(1-x_i y_i)^{-1}
\prod_{i,j=1}^n(1-x_i y_j)^{-\beta}
=
\sum_{\eta\in\Lambda_n}
q_\eta(x)y^\eta.
\ee
The (combinatorial) inner product
$\langle\bullet,\bullet\rangle_{n}^{\rm c}$ 
is then defined by 
$\langle q_\nu,x^\eta\rangle_{n}^{\rm c}=\delta_{\nu\eta}$ 
\cite{Dunkl1}.
The Jack polynomials with the prescribed symmetry
are orthogonal with respect to the inner product
$\langle\bullet,\bullet\rangle_{n}^{\rm c}$ \cite{BF1}.
Using the norm formula
$(||E_\eta||_{n}^{\rm c})^2
=\langle E_\eta,E_\eta\rangle_{n}^{\rm c}
=d'_\eta/d_\eta$ for $\eta\in \Lambda_n$
\cite{Sahi}
and the transformation properties (\ref{trans-property})
of the non-symmetric Jack polynomials,
we can prove
\be
\label{c-norm}
(||K_\mu||_{n}^{\rm c})^2
=
\langle K_\mu,K_\mu\rangle_{n}^{\rm c}
=
\frac{n_\downarrow ! n_\uparrow !}{\rho_\mu}
\frac{d'_\mu}{d_\mu}
\ee
for $\mu \in \Lambda^{\downarrow\uparrow}_n$. 

\noindent
b) integral norm:
For functions $f(x)$ and $g(x)$ in complex variables
$x=(x_1,\cdots,x_n)$, 
we define the inner product 
$\langle \bullet,\bullet\rangle_{n}^0$ by 
the following formula:
\be
\label{def-norm}
\langle f,g\rangle_{n}^0
=
\prod_{i=1}^n\oint_{|x_i|=1}\frac{{\rm d}x_i}{2\pi\mbox{i}x_i}
f(x)\overline{g(x)}
|\Delta(x)|^{2\beta}. 
\ee
Here 
$\Delta(x)
=\prod_{i,j\in I
       \atop
       {\scriptscriptstyle i<j}}
(x_i-x_j)$
is the van der Monde determinant and 
$\overline{g(x)}$ denotes the complex conjugation of $g(x)$.
The Jack polynomials with the prescribed symmetry
are orthogonal with respect to the inner product
$\langle\bullet,\bullet\rangle_{n}^0$ \cite{BF1}.
It is known \cite{BF2} 
that 
\begin{equation}
\langle E_\eta,E_\eta\rangle_{n}^0
/\langle E_\eta,E_\eta\rangle_{n}^{\rm c}
=\frac{\Gamma(n\beta+1)}
      {\Gamma(\beta+1)^n}\frac{e_\eta}{e'_\eta} 
\label{Ei-norm}
\end{equation} 
for $\eta \in \Lambda_n$. 
We note that the right-hand side of (\ref{Ei-norm}) 
depends on $\eta$ through $\eta^+$. 
Thus the relation (\ref{Ei-norm}) immediately leads to 
 \begin{equation}
\langle K_\mu,K_\mu\rangle_{n}^0
/\langle K_\mu,K_\mu\rangle_{n}^{\rm c}
=\frac{\Gamma(n\beta+1)}
      {\Gamma(\beta+1)^n}\frac{e_\mu}{e'_\mu}.   
\label{Kmu-norm}
\end{equation}
As a result of (\ref{c-norm}) and (\ref{Kmu-norm}), 
we obtain
\be
\label{i-norm}
(||K_\mu||_{n}^0)^2
=
\langle K_\mu,K_\mu\rangle_{n}^0
=
\frac{n_\downarrow ! n_\uparrow !}
     {\rho_\mu}
\frac{\Gamma(n\beta+1)}{\Gamma(\beta+1)^n}
\frac{e_\mu d'_\mu}{e'_\mu d_\mu}
\ee
for $\mu \in \Lambda^{\downarrow\uparrow}_n$.

\noindent
c) Cauchy formula:
For the coordinate $x=(x_1,\cdots,x_n)$
and a fixed integer $m\in I$,
we define 
$x^\downarrow=(x_1,\cdots,x_m)$ and
$x^\uparrow=(x_{m+1},\cdots,x_n)$.
The Cauchy formula for the Jack polynomials 
with the prescribed symmetry
is given by
\ba
\label{Cauchy}
\prod_{s=\downarrow,\uparrow}
\prod_{i,j\in I_s\,}
(1-x_i y_j)^{-1}
\prod_{i,j\in I}
(1-x_i y_j)^{-\beta}
=
n_\downarrow ! n_\uparrow !
\sum_{\mu\in \Lambda_n^{\downarrow\uparrow}}
\left(||K_\mu||_n^{\rm c}\right)^{-2}
\tilde K_\mu(x) \tilde K_\mu(y), 
\ea
where 
$\tilde K_\mu(x)=K_\mu(x)
/(\Delta(x^\downarrow)\Delta(x^\uparrow))$
with  
$\Delta(x^s)$ = $\prod_{i,j\in I_s
       \atop
       {\scriptscriptstyle i<j}}
(x_i-x_j)$, $(s=\downarrow,\uparrow)$.
The proof of the Cauchy formula (\ref{Cauchy}) 
is based on the Cauchy formula for the non-symmetric
polynomials 
$\Omega(x|y)
=\sum_{\eta\in \Lambda_n}\left(||E_\eta||_n^{\rm c}\right)^{-2}
E_\eta(x)E_\eta(y)$ 
\cite{Sahi}
and the Cauchy's determinant
identity (see also \cite{BF2}).
The proof also requires the transformation properties
(\ref{trans-property}) and (\ref{coeff}) 
together with those for 
$d_\eta$ and $d'_\eta$ \cite{Sahi}.

\noindent
d) evaluation formula:
Using the evaluation formula for 
the non-symmetric Jack polynomials 
$E_\eta(\underbrace{1,\cdots,1}_{n})=e_\eta/d_{\eta}$
\cite{Sahi}
and new skew operators \cite{Dunkl3},
Dunkl obtained the evaluation formula
for the Jack polynomials
with the prescribed symmetry\footnote{%
The polynomial $E_\eta$ for a composition $\eta$ 
in this paper is different from $\zeta_\eta$ 
in \cite{Dunkl3} by the constant factor $d'_\eta/d_\eta$; 
$E_\eta =(d'_\eta/d_\eta)\zeta_\eta$.}
\be
\label{evaluation}
\tilde K_\mu(\underbrace{1,\cdots,1}_{n})
=
\beta^{-|\delta|}
\frac{1}{e_\delta}
\frac{e_\mu}{d_{\mu}}
\frac{\pi_{\mu}}
     {\rho_{\mu}},  
\ee
where
\begin{equation}
\label{pi}
\pi_\mu
=
\prod_{s=\downarrow,\uparrow}
\prod_{i,j\in I_s
       \atop
       {\scriptscriptstyle i<j}}
\left(\bar{\mu}_i-\bar{\mu}_j-\beta\right) 
\end{equation}
for $\mu \in \Lambda_n ^{\downarrow\uparrow}$. 
In (\ref{evaluation}), 
the composition 
$\delta\in \Lambda_n ^{\downarrow\uparrow}$ 
is introduced as $\delta
=\delta(n^\downarrow,n^\uparrow)
:=(\delta^\downarrow,\delta^\uparrow)$
with $\delta^\downarrow=(n^\downarrow-1,\cdots,1,0)$
and $\delta^\uparrow=(n^\uparrow-1,\cdots,1,0)$. 

\subsection{Binomial formula}
\label{binomial-formula}
In this subsection, 
we derive the binomial formula
with the use of the Cauchy formula (\ref{Cauchy}) 
and evaluation formula (\ref{evaluation}). 
For $a\in\bC$, the binomial formula for
the Jack polynomials with the prescribed symmetry is given by
\be
\label{binomial}
\prod_{s=\downarrow,\uparrow}
\prod_{i\in I_s}
(1-x_i)^{a-n_s}
=
\sum_{\mu\in \Lambda_n^{\downarrow\uparrow}}
\chi_\mu(a)
\tilde K_\mu(x), 
\ee
where
\be
\label{chi}
\chi_\mu(a)
=
\beta^{-|\mu|}
\frac{(1-a)_{\mu^+}}
     {(1-a)_{\delta^+}}
\frac{\pi_\mu}{d'_{\mu}}.
\ee
Here, for an indeterminate $t$, a partition $\lambda$
and a parameter $\beta$,
the generalized shifted factorial is defined by
\be
(t)_\lambda
=
\prod_{i=1}^n
\frac{\Gamma(t-\beta(i-1)+\lambda_i)}{\Gamma(1-\beta(i-1))}.
\ee
We can prove the formula (\ref{binomial}) 
in a way similar to that used in  
Lemma 5.2 in \cite{Sahi} and 
Proposition 2.4 in \cite{BF2}.

Let $n'$ be $n'_\downarrow+n'_\uparrow(\geq n)$ 
for that pair ($n'_\downarrow$, $n'_\uparrow$) 
of positive integers $n'_\downarrow$ and $n'_\uparrow$ 
which satisfies 
$n'_\downarrow-n'_\uparrow=n_\downarrow-n_\uparrow$.
We consider the Cauchy formula (\ref{Cauchy}) 
in $n'$ variables. First we discuss the left-hand side; 
the left-hand side of (\ref{Cauchy}) becomes
\begin{equation}
\prod_{s=\downarrow,\uparrow}
\prod_{i,j\in I'_s}
\left(1-x_i y_j\right)^{-1}
\prod_{i,j\in I'}\left(1-x_i y_j\right)^{-\beta},
\label{lhs}
\end{equation}
with $I'_\downarrow =[1,n'_\downarrow]$,  
$I'_\uparrow =[n'_\downarrow +1,n']$ 
and $I' =[1,n']$. Now we set $x_{n_\downarrow+1}
=\cdots=x_{n'_\downarrow}=0$,
$x_{n'_\downarrow+n_\uparrow+1}
=\cdots=x_{n'_\downarrow+n'_\uparrow}=0$
and
$y_{1}
=\cdots=y_{n'_\downarrow+n'_\uparrow}=1$. 
Further we replace $(x_{n'_\downarrow+1},
\cdots,x_{n'_\downarrow+n_\uparrow})$ by 
$(x_{n_\downarrow+1},
\cdots,x_{n_\downarrow+n_\uparrow})$. 
The expression (\ref{lhs}) then turns into
\begin{equation}
\prod_{i\in I_\downarrow=[1,n_\downarrow]}
(1-x_i)^{-n'\beta-n'_\downarrow}
\prod_{j\in I_\uparrow=[n_\downarrow+1,n]}
(1-x_j)^{-n'\beta-n'_\downarrow+n_\downarrow-n_\uparrow}.
\label{lhs2}
\end{equation}

Next we discuss the right-hand side of the Cauchy formula  
in $n'$ variables. 
We set $y_i=1$ for $1\leq i\leq n'$. 
With the use of the evaluation formula (\ref{evaluation}), 
we immediately see that right-hand side 
of the Cauchy formula becomes
\begin{equation}
\sum_{\nu }
\beta^{-\vert \delta'\vert}
\frac{e_\nu \pi_\nu}{e_{\delta'}d'_\nu} 
\tilde K_\nu(x). 
\label{rhs}
\end{equation}
Here the sum is taken 
over $\nu \in \Lambda_{n'}^{\downarrow\uparrow}$. 
The symbol $\delta'=(\delta'^{\downarrow},\delta'^{\uparrow})$ 
denotes 
the composition 
$\delta(n'_\downarrow,n'_\uparrow)
\in \Lambda_{n'}^{\downarrow\uparrow}$. 

Now we set $x_{n_\downarrow+1}=\cdots=x_{n'_\downarrow}=0$
and 
$x_{n'_\downarrow+n_\uparrow+1}
=\cdots=x_{n'_\downarrow+n'_\uparrow}=0$ in (\ref{rhs}). 
Non-vanishing contributions in the sum (\ref{rhs}) 
then come from only the compositions 
$\nu =(\nu^\downarrow,\nu^\uparrow)
\in \Lambda_{n'}^{\downarrow\uparrow}$ 
satisfying 
$l(\nu^\downarrow-\delta'^\downarrow)\le n_\downarrow$ 
and $l(\nu^\uparrow-\delta'^\uparrow)\le n_\uparrow$
where 
$\nu^s-\delta'^s
=(\nu^s_1-\delta'^{s}_1,\cdots,
\nu^s_{n'_s}-\delta'^{s}_{n'_s})\in\Lambda_{n'_s}^\dagger
$, $(s=\downarrow,\uparrow)$.
The reason is as follows: 
If either 
$l(\nu^\downarrow-\delta'^\downarrow)>n_\downarrow$ 
or 
$l(\nu^\uparrow-\delta'^\uparrow)>n_\uparrow$ 
holds for the composition 
$\nu =(\nu^\downarrow,\nu^\uparrow)
\in \Lambda_{n'}^{\downarrow\uparrow}$, 
then, in each monomial of the polynomials $K_\nu$, 
the minimum power of $x_{n_\downarrow +1}$ 
is 1 or that of $x_{n'_\downarrow+n_\uparrow +1}$ is 1. 
Therefore those $\nu$ do not contribute to the sum 
in (\ref{rhs}) when both $x_{n_\downarrow +1}$ 
and $x_{n'_\downarrow+n_\uparrow +1}$ are set to be zero. 

For a composition 
$\nu =(\nu^\downarrow,\nu^\uparrow)
\in \Lambda_{n'}^{\downarrow\uparrow}$ 
satisfying 
$l(\nu^\downarrow-\delta'^\downarrow)\le n_\downarrow$ 
and  $l(\nu^\uparrow-\delta'^\uparrow)\le n_\uparrow$, 
the composition 
$\mu=(\mu^\downarrow,\mu^\uparrow)
\in\Lambda_n^{\downarrow\uparrow}$ 
can be defined as the composition satisfying the relation 
\begin{equation}
\mu^\downarrow-\delta^\downarrow
=\nu^\downarrow-\delta'^\downarrow
$ and $
\mu^\uparrow-\delta^\uparrow
=\nu^\uparrow-\delta'^\uparrow. 
\label{relation}
\end{equation}
When the relation (\ref{relation}) holds 
for compositions $\mu \in \Lambda_n^{\downarrow \uparrow}$ 
and $\nu \in \Lambda_{n'}^{\downarrow \uparrow}$, 
we can then find the following consequences: 
First, the two polynomials 
\begin{equation}
\tilde K_\nu(x_1,\cdots,x_{n_\downarrow},
\underbrace{0,\cdots,0}_{n'_\downarrow-n_\downarrow},
x_{n'_\downarrow+1},\cdots,x_{n'_\downarrow+n_\uparrow},
\underbrace{0,\cdots,0}_{n'_\uparrow-n_\uparrow})
\end{equation} 
and
\begin{equation}
\tilde K_\mu(x_1,\cdots,x_{n_\downarrow},
x_{n'_\downarrow+1},\cdots,x_{n'_\downarrow+n_\uparrow})
\end{equation}
are equal. From now on, 
we replace
$x_{n'_\downarrow+1},\cdots,x_{n'_\downarrow+n_\uparrow}$ by
$x_{n_\downarrow+1},\cdots,x_n$. 
Second,
$\pi_\nu/d'_\nu
=\beta^{\vert\nu\vert-\vert\mu\vert}
\pi_\mu/d'_\mu$.
Third, the expression $e_{\nu}/e_{\delta'}$ 
can be rewritten as
\begin{eqnarray}
& &
\frac{\prod_{s\in \nu }((a'(s)+1)/\beta +n'-l'(s))}
     {\prod_{s\in \delta '}((a'(s)+1)/\beta +n'-l'(s))}
\nonumber\\
&=&
\frac{\prod_{s\in \mu }
((a'(s)+n'_\downarrow-n_\downarrow+1)/\beta +n'-l'(s))}
{\prod_{s\in \delta }
((a'(s)+n'_\downarrow-n_\downarrow+1)/\beta +n'-l'(s))}
\nonumber\\
&=&
\beta^{\vert\delta\vert-\vert\mu\vert}
\frac{(1+n'\beta +n'_\downarrow-n_\downarrow)_{\mu^+}}
     {(1+n'\beta +n'_\downarrow-n_\downarrow)_{\delta^+}}.
\label{ee}
\end{eqnarray}
(Notice the relation
$\prod_{s\in \eta}
((a'(s)+k)/\beta+k'-l'(s))
=\beta^{-|\eta|}(k'\beta+k)_{\eta^\dagger}$
for 
$\eta \in \Lambda_{n}$ and integers $j, k$.) 
As a result of these three relations, 
the expression (\ref{rhs}) can be rewritten as
\begin{equation}
\sum_{\mu }
\beta^{-\vert \mu\vert}
\frac{(1+n'\beta +n'_\downarrow-n_\downarrow)_{\mu^+}}
     {(1+n'\beta +n'_\downarrow-n_\downarrow)_{\delta^+}}
\frac{\pi_\mu}{d'_\mu} 
\tilde K_\mu(x),  
\label{rhs2}
\end{equation}
where the sum is taken over 
$\mu \in \Lambda_{n}^{\downarrow\uparrow}$. 

Now we obtain the relation 
\begin{equation}
\prod_{i\in I_\downarrow}
(1-x_i)^{-n'\beta-n'_\downarrow}
\prod_{j\in I_\uparrow}
(1-x_j)^{-n'\beta-n'_\downarrow+n_\downarrow-n_\uparrow}
=\sum_{\mu }
\beta^{-\vert \mu\vert}
\frac{(1+n'\beta +n'_\downarrow-n_\downarrow)_{\mu^+}}
     {(1+n'\beta +n'_\downarrow-n_\downarrow)_{\delta^+}}
\frac{\pi_\mu}{d'_\mu} 
\tilde K_\mu(x),   
\label{binomial-proof}
\end{equation}
from the expressions (\ref{lhs2}) and (\ref{rhs2}).
We notice that the right-hand side of (\ref{binomial-proof}) 
is a polynomial of $n'\beta +n'_\downarrow$ 
and hence the relation (\ref{binomial-proof}) 
also holds for arbitrary complex values 
of $n'\beta +n'_\downarrow$. 
Further we replace $-n'\beta -n'_\downarrow+n_\downarrow$ 
by a complex variable $a$. 
Consequently we obtain the binomial formula (\ref{binomial}).

\section{Hole propagator for the
spin Calogero-Sutherland model}
\label{propagator}
\subsection{Result for a finite number of particles}
\label{finite}
In this section, for arbitrary non-negative integer $\beta$, 
we exactly compute the hole propagator
of the SU(2) spin CS model with a finite number of particles.

We consider the $2M$-particle system whose Hamiltonian 
is given by 
(\ref{hamiltonian}) with $p=2$.
We assume that $M$ is an odd integer.
The hole propagator of the model is given by
\begin{equation}
\label{def-propagator}
G\left(r,t\right)\nonumber\\
= 
{}_{2M}\langle 0\vert
\hat \psi_\uparrow ^{\dagger}\left(r,t\right)
\hat\psi_\uparrow\left(0,0\right)
\vert 0\rangle_{2M}/
{}_{2M}\langle 0\vert 0\rangle_{2M},
\end{equation}
where 
$\vert 0\rangle_{2M}$ represents the singlet ground state 
for $2M$-particle system. 
The operator $\hat\psi_\uparrow(r,t)
=\exp({\rm i} \hat{H}_{2M-1}t)
\hat\psi_\uparrow(r)\exp(-{\rm i}\hat{H}_{2M}t)$ 
is the Heisenberg representation of 
the annihilation operator $\hat\psi_\uparrow(r)$
of particles with spin up 
which acts on the $2M$-particle states.

In the following, the statistics of particles 
are chosen as boson (fermion) 
for odd (even) $\beta$, so that
we can set $P_{ij}=\left(-1\right)^{\beta +1} s_{ij}$. 
(Notice that $P_{ij}s_{ij}$ 
is nothing but the particle exchange operator for a pair 
$(i,j$).)

Along the calculations in \cite{Green}, 
the hole propagator reduces to the following expression:
\begin{equation}
\label{tildebeta}
G(r,t)
=
c_0 
\prod_{i=1}^{2M-1}
\oint_{\vert x_i\vert=1}\frac{{\rm d}x_i}{2\pi{\rm i} x_i}
\bar{\Delta}^\beta(x)\bar{\Theta}(x;\beta)
{\rm exp}
\left[-{\rm i}\left(\tilde {\cal H}-E_{2M}^0\right)t
+{\rm i}\tilde {\cal P} r\right]
\Delta^\beta(x)\Theta(x;\beta),
\end{equation}
where complex coordinates $x=(x_1,\cdots,x_{2M-1})$
are related to original coordinates $X$ in (\ref{hamiltonian})
by the formulae
$x_i = \exp(2\pi {\rm i} X_i/L)$ for $1\leq i\leq 2M-1$.
The constant factor $c_0$ is given by $\rho_0/(2D(M))$, 
in terms of
the mean density of particles $\rho_0=2M/L$ and 
\begin{eqnarray}
\label{2M-D}
D(M)
&=&
\prod_{i=1}^{2M}
\oint_{|z_i|=1}\frac{{\rm d}z_i}{2\pi\mbox{i}z_i}
\prod_{1\leq i<j\leq 2M}
|z_i-z_j|^{2\beta}
\prod_{1\leq i<j\leq M}
|z_i-z_j|^{2}
\prod_{M+1\leq i<j\leq 2M}
|z_i-z_j|^{2}
\nonumber\\
&=&
\frac{M!}{(2\beta+1)^M}
\frac{\Gamma((2\beta+1)M+1)}{\Gamma(\beta+1)^{2M}}.
\end{eqnarray}
In (\ref{tildebeta}), 
the function $\Theta(x;\beta )$ has the form	
\be
\label{theta}
\Theta(x;\beta )
=
\prod_{i=1}^M(1-x_i)^{\beta}
\prod_{i=M+1}^{2M-1}(1-x_i)^{\beta +1}
\prod_{1\leq i<j\leq M}(x_i-x_j)
\prod_{M+1\leq i<j\leq 2M-1}(x_i-x_j). 
\ee
Further the symbols $\tilde {\cal H}$, $E_{2M}^0$ 
and $\tilde {\cal P}$ denote the Hamiltonian, 
the ground state energy of $2M$-particle system 
and total momentum, respectively. 
In terms of the complex variables $x$, 
the expressions for  $\tilde {\cal H}$ 
and $\tilde {\cal P}$ are given by
\ba
\label{reduced-hamiltonian}
\tilde {\cal H}
&=&
\left(\frac{2\pi}{L}\right)^2
\left[\sum_{i=1}^{2M-1}
\left(x_i \frac{\partial }{\partial x_i}
-\Delta P\right)^2
-
\sum_{1\leq i<j\leq 2M-1}
\frac{
2\beta\left(\beta -\left(-1\right)^\beta s_{ij}\right)x_i x_j}
{\left(x_i -x_j\right)^2}\right],
\\
\label{reduced-momentum}
\tilde {\cal P}
&=&
\frac{2\pi}{L}
\sum_{i=1}^{2M-1}
\left(x_i \frac{\partial }{\partial x_i}
-\Delta P\right),
\ea
respectively. Here $ \Delta P$ 
denotes $(\beta(2M-1)+M-1)/2$.

Now we introduce a transformed Hamiltonian 
$\hat {\cal H}$ and momentum $\hat {\cal P}$ as 
\ba
\label{gauged-hamiltonian}
\hat {\cal H}
&=&
\Delta^{-\beta }
\tilde {\cal H}\Delta^\beta -E_{2M}^0,
\\
\label{gauged-momentum}
\hat {\cal P}
&=&
\Delta^{-\beta }
\tilde {\cal P}\Delta^\beta.
\ea
Using the notations in the section ref{math-pre}
with $n=2M-1,\,n_\downarrow=M$, and $n_\uparrow=M-1$,
the expression (\ref{tildebeta}) turns into 
\be
\label{propagator-by-i-norm}
G(r,t)
=
c_0
\langle
\Theta, 
{\rm exp}
\left(-{\rm i}\hat {\cal H}t+{\rm i}\hat {\cal P} r\right)
\Theta
\rangle_{2M-1}^{0}.
\ee

In (\ref{propagator-by-i-norm}), 
our problem has reduced to the spectral decomposition of
$\Theta(x;\beta)$ 
in terms of the joint eigenfunctions of $\hat {\cal H}$
and $\hat {\cal P}$. From the following two observations, 
we can see 
that the Jack polynomials with 
the prescribed symmetry are proper bases of the decomposition: 
First, both $\Theta(x)$ and $K_\mu(x)$ 
are polynomials with a common 
symmetric property; 
$s_{i}\Theta(x)=-\Theta(x)$  
and $s_{i}K_\mu(x)=-K_\mu(x)$ 
for $i\in I_\downarrow\backslash\left\{n_\downarrow\right\}$ 
or $\in I_\uparrow\backslash\left\{n\right\}$. 
Second, the polynomials $K_\mu$ 
are joint eigenfunctions of $\tilde {\cal H}$ 
and $\tilde {\cal P}$ with the eigenvalues  
\ba
\label{energy-finite}
\omega(\mu)
&=&\left(\frac{2\pi}{L}\right)^2
\sum_{i=1}^{2M-1}
\left(\bar \mu_i -\frac{M-1}{2}+\beta\right)^2,
\\
\label{momentum-finite}
q(\mu)
&=&
\frac{2\pi}{L}\sum_{i=1}^{2M-1}\left(\mu_i
-\frac{\beta(2M-1)+M-1}{2}\right).  
\ea
respectively.  In the following paragraphs, 
we discuss the second issue. 

{}From the expression (\ref{bareta}), 
we can see that $E_\eta(x)$ for $\eta \in \Lambda_n$ 
and $x=(x_1,\cdots,x_n)$ 
is an eigenfunction of the operators 
$\sum_{i=1}^{n}\hat d_i^k$ 
with the eigenvalue 
$\sum_{i=1}^n \bar \eta_i^k$ for $k=1,\cdots,n$.
Further the operators $\sum_{i=1}^n\hat d_i^k$ 
for $k=1,\cdots,n$ commute with permutations 
$\sigma\in S_{n_\downarrow}\times S_{n_\uparrow}$. 
In addition, $K_\mu$ can be constructed 
by a `symmetrization' of $E_\mu$ 
(see the expression (\ref{KetaEeta})). 
Therefore the Jack polynomials with the prescribed symmetry 
$K_\mu$ for $\mu\in \Lambda_n^{\downarrow\uparrow}$ 
are eigenfunctions of $\sum_{i=1}^n\hat d_i^k$   
with the eigenvalue 
$\sum_{i=1}^n \bar \mu_i^k$ for $k=1,\cdots,n$.

On the other hand, in terms of the Cherednik-Dunkl operators 
(\ref{dunkl}), 
the Hamiltonian $\hat{\cal H}$ 
and total momentum $\hat{\cal P}$ can be written as
\begin{equation}
\hat{\cal H}
=\left(\frac{2\pi}{L}\right)^2\sum_{i=1}^{2M-1}
\left(\hat d_i -\frac{M-1}{2}+\beta\right)^2,
\label{Hamil-by-dunkl}
\end{equation}
and
\begin{equation}
\hat{\cal P}
=\frac{2\pi}{L}
\left[
\sum_{i=1}^{2M-1}
\left(\hat d_i -\frac{M-1}{2}\right)
+\frac{\beta\left(2M-1\right)}{2}
\right].
\label{momentum-by-dunkl}
\end{equation}
These expressions lead the fact that the joint eigenfunctions 
of $\sum_{i=1}^{2M-1}\hat d_i^k $ for $k=1,2$
are those of $\hat{\cal H}$ 
and $\hat{\cal P}$. From this fact, 
we find that the polynomials $K_\mu$ 
are joint eigenfunctions of $\hat{\cal H}$ and $\hat{\cal P}$. 
The eigenvalues (\ref{energy-finite}) 
and (\ref{momentum-finite}) follow from (\ref{bareta}), 
(\ref{Hamil-by-dunkl}) and (\ref{momentum-by-dunkl}). 

We notice that the binomial formula (\ref{binomial}) 
is useful in rewriting (\ref{propagator-by-i-norm}), 
because the spectral decomposition of $\Theta(x;\beta)$ 
in terms of $K_\mu(x)$ 
is a special case of the formula (\ref{binomial}). 
Now we can express $\Theta$ in terms of $K_\mu$ as
\begin{equation}
\label{expansion}
\Theta(x;\beta)
=
\sum_{\mu \in \Lambda_{2M-1}^{\downarrow\uparrow}}
\chi_\mu(\beta+M) K_\mu(x;\beta),
\end{equation}
with the coefficients $\chi_\mu(\beta+M)$ (\ref{chi}). 

Using the orthogonal properties of 
the Jack polynomials with the prescribed symmetry and 
the spectral decomposition (\ref{expansion}),
we have
\begin{equation}
G(r,t)
=
\frac{\rho_0}{2} 
\sum_{\mu \in \Lambda_{2M-1}^{\downarrow\uparrow}}
\left(\chi_\mu (\beta+M)\right)^2 
\left(
\frac{||K_\mu||_{2M-1}^0}{||K_{\delta(M,M)}||_{2M}^0}
\right)^2
\exp
\left[-{\rm i}\left(\omega(\mu)t-q(\mu) r\right)\right].
\label{hole-finite}
\end{equation}
(Notice that 
$K_{\delta(M,M)}(x)=\Delta(x^\downarrow)\Delta(x^\uparrow)$
and then 
$D(M)=(||K_{\delta(M,M)}||_{2M}^0)^2$.)
Every factor in (\ref{hole-finite}) 
is available from the expressions
(\ref{i-norm}), (\ref{chi}), 
(\ref{energy-finite}) and (\ref{momentum-finite}). 
The expression (\ref{hole-finite}) 
is our main result in this subsection.

It is important to consider 
the condition that 
intermediate states $\mu$ can contribute to the sum 
in (\ref{hole-finite}).  
By a close examination of the expression (\ref{chi}), 
we find the relation:
\ba
\label{selection-rule}
\chi_\mu (\beta+M)\ne 0
&\Leftrightarrow&
(1,\beta+M)\notin\mu^+.
\ea
This relation leads a selection rule; 
only those states $\mu$ the largest entry of 
which is less than $\beta+M$ contribute to the sum 
in (\ref{hole-finite}). 
Here we remark on the parametrization of 
the relevant intermediate state.
For a composition 
$\mu\in \Lambda_n^{\downarrow\uparrow}$,
there is a unique composition 
$\hat{\mu}=(\hat{\mu}^\downarrow,\hat{\mu}^\uparrow)
\in \Lambda_n$ such that 
$\mu=\hat{\mu}+\delta$.
Notice that 
$\hat{\mu}^s \in\Lambda_{n_s}^\dagger$ 
for $s=\downarrow,\uparrow$.
Using these notations, we have 
\ba
\label{selection-rule2}
\chi_\mu (\beta+M)\ne 0
&\Leftrightarrow&
(1,\beta+1)\notin\hat{\mu}^\downarrow
\,\mbox{and}\,
(1,\beta+2)\notin\hat{\mu}^\uparrow,
\ea
where 
$\hat{\mu}^\downarrow$ and $\hat{\mu}^\uparrow$
are regarded as partitions.
In the decomposition $\mu=\hat{\mu}+\delta$, 
$\delta$ represents the condensate or the pseudo Fermi sea
and $\hat{\mu}$ represents the excitations. From 
above conditions, we see that such partitions
$\hat{\mu}^\downarrow$ and $\hat{\mu}^\uparrow$
are parametrized by $\beta$ and $\beta+1$
integers, respectively.
Therefore, 
the relevant intermediate states $\mu$ 
can be parametrized by $2\beta +1$ integers.

An example of those states is shown in Fig. 1, 
where we take $\beta=2$, $M=7$, 
and $\mu=(2,2,1,1,0,0,0,3,3,2,2,1,0)
+\delta(n_\downarrow=7,n_\uparrow=6)$. 
The open blocks correspond to 
$\delta(n_\downarrow=7,n_\uparrow=6)$ and 
the shaded ones represent the excitations. 

In next subsection, 
we will see that another but equivalent
parametrization leads 
$(2\beta +1)$-fold integral representation 
for $G(r,t)$ in the thermodynamic limit. 

\subsection{Thermodynamic limit}
\label{thermo-lim}
In this subsection, we derive the expression 
(\ref{conjecture}) for the hole propagator 
in the thermodynamic limit.
First of all, to take the thermodynamic limit,
we parametrize the intermediate states in the following way.
Let $\mu=(\mu^\downarrow,\mu^\uparrow)
\in \Lambda_{2M-1}^{\downarrow\uparrow}$ 
be a composition which satisfies the condition 
(\ref{selection-rule}). From 
the consideration on the intermediate states,
the complementary set of 
$\{\mu_i^\downarrow\}_{i=1}^M$ in
$\left\{M+\beta-1,M+\beta-2,\cdots,1,0\right\}$ 
is well-defined.
We define $\left\{p_k\right\}_{k=1}^\beta$ 
so that $p_1>\cdots>p_{\beta}$ 
and $\left\{M+\beta-1,M+\beta-2,\cdots,1,0\right\}
\setminus \{\mu_i^\downarrow\}_{i=1}^M
=\left\{M+\beta-1-p_k\right\}_{k=1}^{\beta}$. 
Similarly, we define $\left\{q_l\right\}_{l=1}^{\beta +1}$ 
so that $q_{1}>\cdots>q_{\beta+1}$
and $\left\{M+\beta-1,M+\beta-2,\cdots,1,0\right\} 
\setminus \{\mu_i^\uparrow\}_{i=1}^{M-1}
=\left\{M+\beta-1-q_l\right\}_{l=1}^{\beta +1}$. 
The resultant set of 
$\left\{p_k,q_l\right\}$ 
exhausts the relevant intermediate state in 
(\ref{hole-finite}). In Fig. 1, the new parameters 
$\left\{p_1,p_2,q_1,q_2,q_3\right\}$ are also shown. 

Furthermore we shall introduce a set of notations.  
We respectively define 
$\tilde p_k\,(1\leq k\leq \beta)$ 
and 
$\tilde q_l\,(1\leq l\leq \beta+1)$
by
\ba
\tilde p_k
&=&
p_k
-\gamma\beta 
\left(
\beta-k+
\sharp
\left\{l'\in\{1,\cdots,\beta+1\}|q_{l'}<p_k
\right\}
\right),\\
\tilde q_l
&=&
q_l
-\gamma\beta
\left(
\beta+1-l
+\sharp\left\{k'\in\{1,\cdots,\beta\}|p_{k'}\le q_l\right\}
\right), 
\ea
where $\gamma=1/(2\beta+1)$. 
We can regard $\tilde p_k $ and $\tilde q_l$ 
as `rapidities' of quasiholes with spin up and down. 
Also, $\bar \Delta_k\,(1\leq k\leq \beta)$ and 
$\hat \Delta_l\,(1\leq l\leq \beta+1)$
are defined as 
$\gamma\beta\sharp
\left\{l'\in \{1,\cdots,\beta+1\}|q_{l'}=p_k -1\right\}$
and
$\gamma\beta\sharp
\left\{k'\in \{1,\cdots,\beta\}|p_{k'}=q_l -1\right\}$,
respectively.
Moreover we introduce 
$U_{kl}=\tilde p_k -\tilde p_l-\bar \Delta_l $ 
for $1\le k<l\le \beta$ 
and $V_{kl}=\tilde q_k -\tilde q_l -\hat \Delta_l$ 
for $1\le k<l \le \beta +1$. 
They describe the interplay between the quasiholes 
with a same spin.
In order to describe the interaction 
between the quasiholes with an opposite spin, 
we introduce $W_{kl}$ for $1\leq k\leq\beta$ 
and $1\leq l\leq \beta+1$ as
\begin{equation}
W_{kl}=\left\{
\begin{array}{cl}
\tilde p_k -\tilde q_l 
-\gamma\left(\beta+1\right)
\sharp\left\{k'\in \{1,\cdots,\beta\}|p_{k'}=q_l +1\right\}
&,\quad\mbox{for }q_l +2\le p_k,\\
1
&,\quad\mbox{for }q_l\le p_k \le q_l +1,\\
\tilde q_l -\tilde p_k +1-\gamma\left(\beta+1\right)
\sharp\left\{l'\in \{1,\cdots,\beta+1\}|q_{l'}=p_k\right\} 
&,\quad\mbox{for }p_k \le q_l-1,
\end{array}\right.
\end{equation}
and
\be
\tilde \Delta_{kl}=
\left\{
\begin{array}{cl}
\gamma\beta,
&\quad\mbox{for }p_k=q_l\mbox{ or }p_k=q_l+1,
\\
0,&\quad\mbox{otherwise}.
\end{array}\right.
\ee
Here $\tilde \Delta_{kl}$ is introduced 
to describe exceptional configurations of rapidities. 

Using these new notations, 
we can rewrite the norm and matrix element more explicitly. 
The expression for 
$\chi_\mu (\beta+M)$ in the new notation is given by
\begin{eqnarray}
\chi_\mu(\beta+M)
&=&
\left(-1
\right)^{%
\sum_{k=1}^\beta p_k+\sum_{l=1}^{\beta+1} q_l -\beta^2}
\prod_{k=1}^{2\beta+1}
\Gamma\big(\gamma k\big)^{-1}
\nonumber\\
&\times&
\prod_{k=1}^\beta 
\Gamma\big(\gamma\left(\beta+1\right)-\bar \Delta_k \big)
\prod_{l=1}^{\beta+1}
\Gamma\big(\gamma\left(\beta+1\right)-\hat \Delta_l \big) 
\nonumber\\
&\times&
\prod_{1\le k< l\le \beta}
\frac{\Gamma\big(U_{kl}+\gamma\left(\beta+1\right)\big)}
     {\Gamma\big(U_{kl}\big)} 
\prod_{1\le k<l \le \beta+1}
\frac{\Gamma\big(V_{kl}+\gamma\left(\beta+1\right)\big)}
     {\Gamma\big(V_{kl}\big)}
\nonumber\\
&\times&
\prod_{k=1}^{\beta}
\prod_{l=1}^{\beta+1}
\frac{\Gamma\big(W_{kl}+\tilde \Delta_{kl}-\gamma\beta\big)}
     {\Gamma\big(W_{kl}\big)}\label{chi-new}.
\end{eqnarray} 
The expression for the norm is given by 
\begin{eqnarray}
\left(
\frac{||K_\mu||_{2M-1}^0}{||K_{\delta(M,M)}||_{2M}^0}
\right)^2
&=&
\frac{\Gamma\big(\beta+1\big)}{M}
\frac{\Gamma\big(M\left(2\beta+1\right)-\beta\big)}
     {\Gamma\big(M\left(2\beta+1\right)\big)}
\prod_{k=1}^{2\beta+1}
\frac{\Gamma\big(M+1-\gamma k\big)}
     {\Gamma\big(M+1-\gamma k-\gamma\beta\big)}
\nonumber\\
&\times&
\prod_{k=1}^{\beta}
\frac{\Gamma\big(1-\bar \Delta_{k}\big)
      \Gamma\big(\tilde p_k+\gamma\left(\beta+1\right)\big)
      \Gamma\big(M-\tilde p_k-\gamma\beta\big)}
     {\Gamma\big(\gamma\left(\beta+1\right)-\bar \Delta_{k}\big)
      \Gamma\big(\tilde p_k +1\big)
      \Gamma\big(M-\tilde p_k\big)}
\nonumber\\
&\times&
\prod_{l=1}^{\beta+1}
\frac{\Gamma\big(1-\hat \Delta_{l}\big)
      \Gamma\big(\tilde q_l+\gamma\left(\beta+1\right)\big)
      \Gamma\big(M-\tilde q_l-\gamma\beta\big)}
     {\Gamma\big(\gamma\left(\beta+1\right)-\hat \Delta_{l}\big)
      \Gamma\big(\tilde q_l +1\big)
      \Gamma\big(M-\tilde q_l\big)}
\nonumber\\
&\times&
\prod_{1\le k<l \le \beta}
\frac{\Gamma\big(U_{kl}+1\big)
      \Gamma\big(U_{kl}\big)}
     {\Gamma\big(U_{kl}+\gamma\left(\beta+1\right)\big)
      \Gamma\big(U_{kl}+\gamma\beta\big)}
\nonumber\\
&\times&
\prod_{1\le k<l \le \beta+1}
\frac{\Gamma\big(V_{kl}+1\big)
      \Gamma\big(V_{kl}\big)}
     {\Gamma\big(V_{kl}+\gamma\left(\beta+1\right)\big)
      \Gamma\big(V_{kl}+\gamma\beta\big)}
\nonumber\\
&\times&
\prod_{k=1}^{\beta}
\prod_{l=1}^{\beta+1}
\frac{\Gamma\big(W_{kl}\big)
      \Gamma\big(W_{kl}+\tilde \Delta_{kl}\big)}
     {\Gamma\big(W_{kl}+\gamma\beta\big)
      \Gamma\big(W_{kl}+\tilde \Delta_{kl}-\gamma\beta\big)}
\label{norm-new}.
\end{eqnarray}

Now we consider the thermodynamic limit, i.e.,
$M\rightarrow\infty$, $L\rightarrow\infty$ with 
$\rho_0=2M/L$ fixed. In this limit, only the configurations
with $p_k$, $q_l$, $\vert p_k-p_l\vert$, $\vert q_k-q_l\vert$ 
and $\vert p_k-q_l\vert$ $\sim {\cal O}(M)$ 
give finite contributions to the hole propagator. 
Let us introduce the normalized velocities 
$u_k$ and $v_l$ of the holes
with up and down spin, respectively. 
These are defined by
\ba
\lim_{M\rightarrow \infty}\frac{2p_k}{M}&=&1-u_k,
\\
\lim_{M\rightarrow \infty}\frac{2q_l}{M}&=&1-v_l.
\ea
In the thermodynamic limit, we can use the Stirling formula:
$\Gamma(z+1)
\rightarrow 
\sqrt{2\pi}z^{z+1/2}\exp\left(-z\right)$ 
for $|z|\rightarrow \infty$. 
Taking the symmetry of the integrand 
into consideration,
the sum in (\ref{hole-finite})
over $\left\{p_k,q_l\right\}$ 
reduces to the integral over $\left\{u_k,v_l\right\}$ 
as
\be
\sum_{0\le p_\beta<\cdots <p_1 \le M+\beta-1}
\,
\sum_{0\le q_{\beta+1}<\cdots <q_1\le M+\beta-1}
\rightarrow \frac{1}{\beta!\left(\beta+1\right)!}
\left(\frac{M}{2}\right)^{2\beta+1}
\prod_{k=1}^\beta \int_{-1}^1 {\rm d}u_k
\prod_{l=1}^{\beta +1}\int_{-1}^1 {\rm d}v_l. 
\ee
Combining the expressions (\ref{hole-finite}), (\ref{chi-new}) 
and (\ref{norm-new}) with the above limiting procedure, 
we arrive at the final expression (\ref{conjecture}) 
for the hole propagator in the thermodynamic limit. 

\section{Acknowledgements}
The authors would like to thank Y. Kuramoto 
for his useful comments on our manuscript. 
YK is grateful to J. Zittartz 
for his hospitality 
at the Institute for Theoretical physics, 
University of K\"oln. 
This work was partly supported by 
the Sonderforschungsbereich 341
K\"oln-Aachen-J\"ulich. 
TY was supported by 
the Core Research for Evolutional Science and Technology 
(CREST)
program of the Science and Technology Agency of Japan.

\begin{figure}
\caption{A diagram that contributes to the hole propagator. 
This figure corresponds to the case with
$\beta=2$, 
$M=7$ 
and 
$\mu=(2,2,1,1,0,0,0,3,3,2,2,1,0)
+\delta(n_\downarrow=7,n_\uparrow=6)$. 
The open and shaded blocks represent the pseudo Fermi sea 
and the excitation, respectively. 
The parameters $\left\{p_k,q_l\right\}$ adopted in 
\ref{thermo-lim} are also shown. }
\end{figure}


\begin{references}
%
%
\bibitem[*]{permanent}
Permanent Address: 
Department of Applied Physics, 
University of Tokyo, Tokyo 113-8656, Japan. 
\bibitem[**]{crest}
CREST researcher:
Japan Science and Technology Corporation (JST),
Kawaguchi 332-0012, Japan.
%
%
%


\bibitem{Cal}
Calogero F 
1969 {\it J. Math. Phys.} {\bf 10} 2191; 2197 

\bibitem{Suth}
Sutherland B 
1971 {\it Phys. Rev.} A{\bf 4} 2019; 
1972 A{\bf 5} 1372; 
1971 {\it J. Math. Phys.} {\bf 12} 246; 251



\bibitem{SLA} 
Simons B D, Lee P A and Al'tshuler B L 
1993 {\it Phys. Rev. Lett.} {\bf 70} 4122; 
1993 {\it Nucl. Phys.} {\bf B409} 487

\bibitem{Ha}
H
a Z N C 
1994 {\it Phys. Rev. Lett.} {\bf 73} 1574; 
1995 {\bf 74} 620 (errata);
1995 {\it Nucl. Phys.} B{\bf 435} 604 

\bibitem{LPS}
Lesage F, Pasquier V and Serban D 
1995 {\it Nucl. Phys.} B{\bf 435} 585


\bibitem{MP}
Minahan J A and Polychronakos A P 
1994 {\it Phys. Rev.} B {\bf 50} 4236


\bibitem{HZ}
Haldane F D M and Zirnbauer M R 
1993 {\it Phys. Rev. Lett.} {\bf 71} 4055

\bibitem{ZH}
Zirnbauer M R and Haldane F D M 
1995 {\it Phys. Rev.} B {\bf 52} 8729


\bibitem{SLP}
Serban D, Lesage F and Pasquier V 
1995 {\it Nucl. Phys}. {\bf B466} 499



\bibitem{HaHaldane}
Ha Z N C and Haldane F D M 
1992 {\it Phys. Rev.} B {\bf 46} 9359

\bibitem{MP2}
Minahan J A and Polychronakos A P 
1993 {\it Phys. Lett.} B {\bf 302} 265


\bibitem{Halperin}
Halperin B I 1983 
{\it Helve. Phys. Acta} {\bf 56} 75 


\bibitem{Green}
Kato Y 
1997 {\it Phys. Rev. Lett.} {\bf 78} 3191


\bibitem{Uglov}
Uglov D 
1998 {\it Commun. Math. Phys.} {\bf 191} 663


\bibitem{Dunkl3}
Dunkl C F 
1997 Orthogonal Polynomials of Type $A$
and $B$ and Related Calogero Models 
{\it preprint} q-alg/9710015


\bibitem{KYA}
Kato Y, Yamamoto T and Arikawa M 
1997 {\it J. Phys. Soc. Jpn.} {\bf 66} 1954




\bibitem{Macd}
Macdonald I G 
1995 {\it Symmetric Functions and Hall Polynomials, 2nd ed.},
(Oxford: Oxford University Press)

\bibitem{KS}
Knop F and Sahi S 
1997 {\it Inv. Math.} {\bf 128} 9

\bibitem{Sahi}
Sahi S 
1996 {\it IMRN} {\bf 20} 997




\bibitem{Opdam}
Opdam E 
1995 {\it Acta. Math.} {\bf 175} 75



\bibitem{Macdonald}
Macdonald I G 
1995 {\it S\'{e}m. Bourbaki} {\bf 797} 1



\bibitem{Dunkl0}
Dunkl C F
1989 {\it Trans. AMS} {\bf 311} 167

\bibitem{Cherednik}
Cherednik I V 
1991 {\it Inv. Math.} {\bf 106} 411


\bibitem{Polychronakos}
Polychronakos A P 
1992 {\it Phys. Rev. Lett.} {\bf 69} 703


\bibitem{BF1}
Baker T H and Forrester P J 
1997 {\it Nucl. Phys.} B {\bf 492} 682


\bibitem{Dunkl1}
Dunkl C F 
Intertwining operators and polynomials
associated with the symmetric group 
{\it Monatsh. Math.} to appear


\bibitem{BF2}
Baker T H and Forrester P J 
1997 Symmetric Jack polynomials from the 
non-symmetric theory 
{\it preprint} q-alg/9707001


%
\end{references}
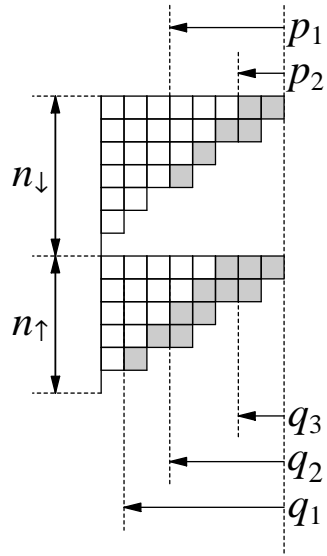
\end{document}